\documentclass[universe,article,accept,moreauthors,pdftex]{Definitions/mdpi}
\usepackage{bm}
\usepackage{bbold}

\newcommand{\GN}{G_\textsc{n}}
\newcommand{\Hu}{\mathcal{H}}
\newcommand{\bnab}{\bm{\nabla}}
\newcommand{\E}{{\mathrm e}}
\newcommand{\I}{{\mathrm i}}
\newcommand{\D}{{\mathrm d}}
\newcommand{\X}{{\mathbf x}}

\renewcommand{\(}{\left(}
\renewcommand{\)}{\right)}
\renewcommand{\[}{\left[}
\renewcommand{\]}{\right]}

\renewcommand{\dot}[1] {\overset{\,_{\mbox{\large .}}}{#1}}

\firstpage{1} 
\pubvolume{8}
\issuenum{1}
\articlenumber{36}
\pubyear{2022}
\copyrightyear{2022}
\externaleditor{\textls[-25]{Academic Editors: Jean-Pierre
Gazeau and Przemyslaw Malkiewicz}}
\datereceived{10 December 2021}
\dateaccepted{6 January 2022}
\datepublished{8 January 2022}
\hreflink{https://doi.org/10.3390/\linebreak universe8010036}

\pdfoutput=1

\Title{Time in Quantum Cosmology}
\TitleCitation{Time in Quantum Cosmology}

\Author{Claus Kiefer $^1$\orcidA{} and Patrick Peter $^{2,}$*\orcidB{}}

\AuthorNames{Claus Kiefer and Patrick Peter}
\AuthorCitation{Kiefer, C.; Peter, P.}
\address{%
$^{1}$ \quad Faculty of Mathematics and Natural Sciences,
Institute for Theoretical Physics, University of Cologne, 
Z\"ulpicher Stra{\ss}e 77a, 50937 Köln, Germany;
kiefer@thp.uni-koeln.de\\

$^{2}$ \quad {${\cal G}\mathbb{R}\varepsilon\mathbb{C}{\cal
O}$}---Institut d'Astrophysique de Paris, CNRS \& Sorbonne
Universit\'e, UMR 7095 98 bis Boulevard Arago, 75014 Paris,
France}

\corres{Correspondence: peter@iap.fr}

\simplesumm{We briefly review the problem of time in quantum 
  cosmology and present two possible solutions.}

\abstract{Time in quantum gravity is not a well-defined
notion despite its central role in the very definition of
dynamics. Using the formalism of quantum geometrodynamics,
we briefly review the problem and illustrate it with two
proposed solutions. Our main application is quantum
cosmology---the application of quantum gravity to the
Universe as a whole.}


\begin{document}

\section{Problem of Time}

The problem of time in quantum
gravity~\cite{Kuchar:1991qf,Isham:1992ms,Halliwell:2002cg,
Kiefer:2012ria, Kiefer:2013jqa,Anderson:2017jij} has been
with us for more than fifty years, having been originally
discussed by Peter Bergmann and his group and by Paul Dirac
in the late 1950s\footnote{The origin of this discussion
can, in fact, be traced back to the pioneering work of
L\'eon Rosenfeld in \oldstylenums{1930}, see the detailed
account by Salisbury in Ref.~\cite{Salisbury:2009cr}.}. It
consists of various issues and questions regarding the
representation of time in both the classical canonical
theory (General Relativity---GR) and its quantization. It is
interesting to note that this ongoing debate on the time
issue can be traced back to Newton and his absolute time and
Leibniz's critique of it (see
Refs.~\cite{Arthur:1985,Barbour:absolute}).  In short, time
in quantum theory is absolute, whereas it is dynamical in
GR.  This ``incompatibility'' a priori renders complicated
the intertwining of both theories into a working quantum
theory of gravity, which is indeed yet to be formulated. We
emphasize that the primary distinction here is between
absolute (background) and {\em dynamical} variables, not so
much between absolute and {\em relative} (e.g.,~in special
relativity) variables; it is the dynamical
variables~\cite{Ehlers:1995} onto which the superposition
principle is being applied in the quantum theory.

So far, no generally accepted theory of quantum gravity
exists. The reasons for this state of affairs are conceptual
and mathematical as well as experimental, because
the effects of quantum gravity are believed to be small in
most situations~\cite{Kiefer:2012ria}. Still, various
approaches exist, from which one may get insights into
aspects of the full theory. One can distinguish between two
broad classes of approaches. In the first class, more or
less heuristic rules are employed to transform a classical
gravitational theory into a corresponding quantum version.
A more proper wording would be ``to guess a theory of
quantum gravity from its classical limit''; this is, in
fact, the method that guided Erwin Schr\"odinger to his
famous wave equation in \oldstylenums{1926}. Such rules can
be applied to any gravity theory, but the most common case
is to apply it to GR. In the second class, attempts are made
to directly construct a fundamental quantum theory (possibly
a unified theory of all interactions) from which quantum
gravity, and finally GR, can be derived. The main example is
the string (or M-) theory, in which quantum gravity
is emergent.

Here, we shall restrict our discussion to the first class,
and moreover, to quantum GR in the Hamiltonian approach with
metric variables (quantum geometrodynamics). This is for two
reasons. First, this approach makes conceptual issues
(notably the problem of time) most transparent. Second, it
is a very conservative approach, at which one
straightforwardly arrives when quantizing GR. Hence, even if
that theory is eventually superseded by a more comprehensive
theory at the most fundamental level (which is likely to be
the case), it can serve as a blueprint for understanding the
concept of time in any quantum gravity theory.

Our paper is organized as follows. In Section~\ref{sec2}, we
review the main features of classical and quantum
geometrodynamics. Section~\ref{sec3} will then focus on the
problem of time and its possible solution at the fundamental
quantum level. There, we will make a careful distinction
between the problem of time at the classical and at the
quantum level. At the classical level, this is also called
{\em background independence} and is not really a
problem---it is a feature of any classical theory that has a
spacetime diffeomorphism invariance. The quantum level is
characterized by the absence of spacetime, in analogy to the
absence of trajectories in quantum mechanics. We review two
possible solutions to the problem of time at the quantum
level. Finally, Section~\ref{sec4} contains a brief summary
and the conclusions.

\section{Classical and Quantum Geometrodynamics \label{sec2}}

\begin{figure}[H]
\centering
\includegraphics[width=11cm]{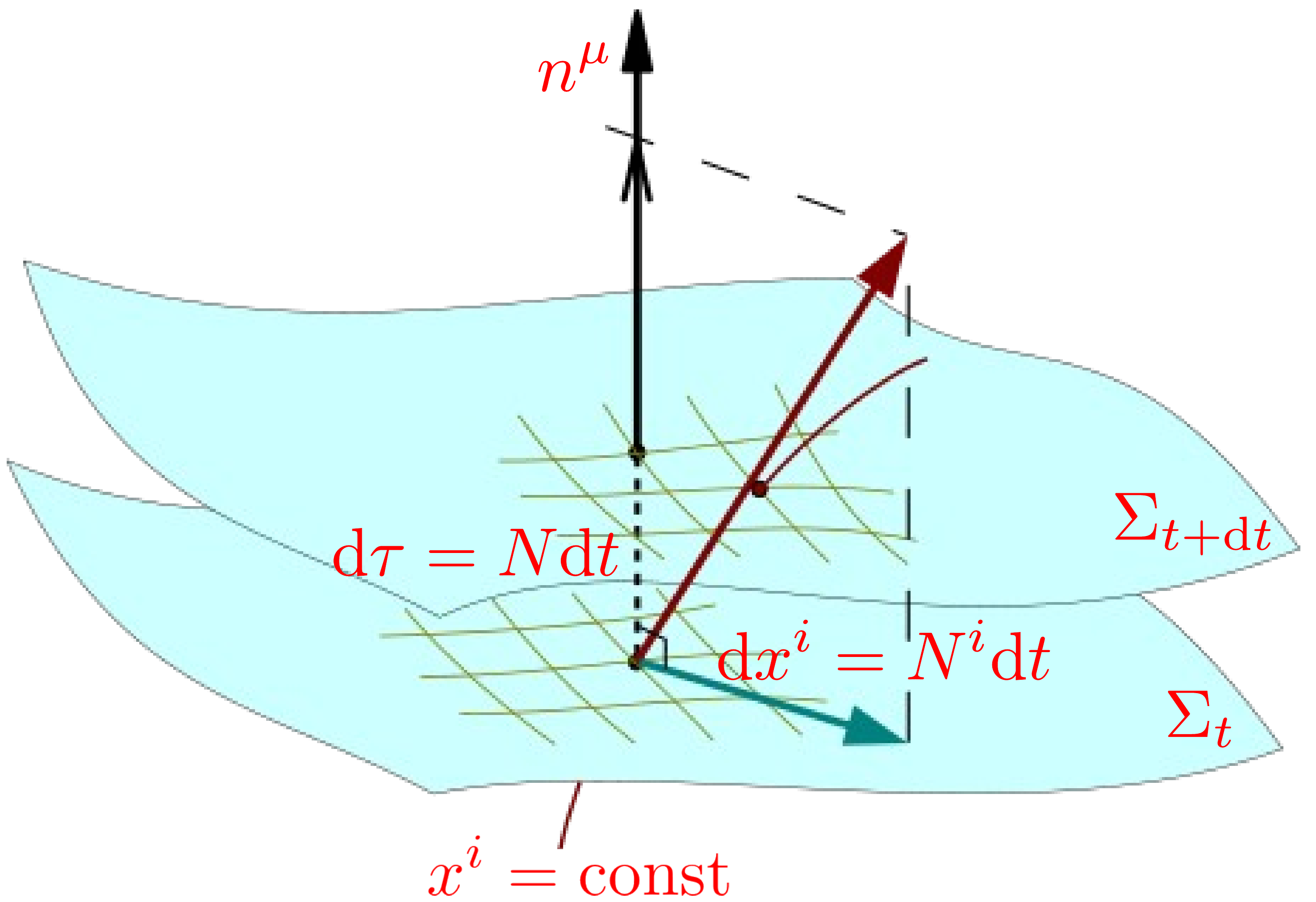}

\caption{Spacetime is split into three-dimensional
hypersurfaces $\Sigma_t$ labeled by a time parameter $t$,
with the orthonormal vector $n^\mu$. The figure illustrates
the various quantities involved in the $3+1$ split of the
metric expansion \eqref{ds2}.}

\label{foliation}
\end{figure}

Quantizing gravity can be performed in a covariant way or
through a formulation of GR in a canonical (Hamiltonian)
form (for example, see Ref.~\cite{Kiefer:2012ria} and the
references therein). The latter approach, on which we will
focus our attention in what follows, begins with a foliation
of spacetime into three-dimensional hypersurfaces, thereby
breaking the manifest four-dimensional diffeomorphism
invariance. This is illustrated in Figure~\ref{foliation},
in which two infinitesimally close leaves $\Sigma_t$ and
$\Sigma_{t+\D t}$ are shown together with the various
relevant geometric quantities.

\unskip   

\subsection{Space and Time Decomposition: The 3 + 1 Split}

In a space $+$ time ($3+1$) decomposition of spacetime,
the general four-dimensional metric $g_{\mu\nu}$ has ten
independent components, which are decomposed into a lapse
function $N$, a shift vector $N^i$, and the
three-dimensional hypersurface- ($\Sigma_t$-) induced metric
$h_{ij}$ (first fundamental form) through the following
equation\footnote{We use units in which $c=1$.}:
\begin{equation}
\D s^2 = g_{\mu\nu} \D x^\mu \D x^\nu = - N^2 \D t^2 +
h_{ij} \( \D x^i + N^i \D t\) \( \D x^j + N^j \D t\),
\label{ds2}
\end{equation}
from which one builds the second fundamental form
\begin{equation}
K_{ij} = - \nabla^{(h)}_j n_i = \frac{1}{2 N} \(
\nabla^{(h)}_j N_i +\nabla^{(h)}_i N_j - \frac{\partial
h_{ij}}{\partial t}\), \label{Kij}
\end{equation}
which is also called the extrinsic curvature of the
hypersurfaces. The relevant action for GR reads, integrated
over the manifold $\mathcal{M}$,as follows:
\begin{equation}
\mathcal{S} = \frac{1}{16\pi\GN} \left[ \int_\mathcal{M}
\!\!\! \sqrt{-g} \ \( R  -2\Lambda \)\D^4 x + 2
\int_{\partial \mathcal{M}} \!\! \!\sqrt{h}\ K \D^3 x
\right] + \mathcal{S}_\mathrm{m} \left[\Phi\(x\)\right],
\label{Sgrav}
\end{equation}
where $K\equiv K^i_{\ i} = h^{ij} K_{ij}$, $h:={\rm det}\
h_{ij}$, and $\GN$ as Newton's constant. This is the
Einstein--Hilbert action, including in the most general
situation a possible cosmological constant $\Lambda$. We
also wrote, for the sake of generality, the Gibbon--Hawking
surface term\footnote{For a recent discussion of boundary
terms, see Ref.~\cite{Feng:2021lfa}.} (integrated over the
boundary $\partial \mathcal{M}$), introduced by Einstein in
\oldstylenums{1916}.  In what follows, however, we shall
restrict attention to compact hypersurfaces $\Sigma_t$,
for which this term identically vanishes.  In \eqref{Sgrav},
we also consider the possibility of a matter component,
encoded as $\mathcal{S}_\mathrm{m}$, with dynamics driven by
generic fields labeled as $\Phi (x)$, which is undetermined
at this stage.

In terms of the $3+1$ decomposition stemming from
\eqref{ds2}, this action can be transformed into its
Arnowitt--Deser--Misner (ADM) \cite{ADM:1962} form given by
the following:
\begin{equation}
\mathcal{S} = \int L\D t = \int \D t \,
\left[\frac{1}{16\pi\GN} \int \D^3 x\, N\sqrt{h} \( K_{ij}
K^{ij} -K^2 +^3\!\! R-2\Lambda \) + L_\mathrm{m} \right],
\label{SKij}
\end{equation}
where $^3\! R$ is the Ricci scalar derived from the
three-dimensional metric $h_{ij}$. The~Lagrangian $L$ thus
defined permits the calculation of the canonical momenta
corresponding to the underlying variables presented above in
\eqref{ds2} and \eqref{Kij}. These are the following:
\begin{equation}
\pi^{ij} := \frac{\delta L}{\delta \dot{h}_{ij}} =
-\frac{\sqrt{h}}{16\pi\GN} \( K^{ij} - h^{ij} K\),
\label{piij}
\end{equation}
for the purely gravitational part, and
\begin{equation}
\pi^{\Phi} := \frac{\delta L}{\delta \dot{\Phi}}  =
-\sqrt{h} \, n^\mu \partial_\mu \Phi = -\frac{\sqrt{h}}{N}
\( \dot{\Phi} - N^i \frac{\partial \Phi}{\partial x^i} \),
\label{piphi}
\end{equation}
in the case of $\Phi(x)$ actually representing a minimally
coupled scalar field; we also used the decomposition $n^\mu
= N^{-1} \left(1, -N^i \right)$ (e.g.,
see~Ref.~\cite{Kiefer:2012ria}). We note that $\GN$
explicitly occurs in~\eqref{piij}, which is why it will
already appear in the {\em vacuum} equation for quantum
gravity (see~\eqref{HPsi} below).

Since neither the shift vector nor the lapse time
derivatives enter the Lagrangian appearing in \eqref{SKij},
their associated momenta are vanishing and merely set the
following primary constraints:
\begin{equation}
\pi^{0} := \frac{\delta L}{\delta \dot{N}} \approx 0 \ \ \ \
\hbox{and} \ \ \ \ \pi^{i} := \frac{\delta L}{\delta
\dot{N^i}} \approx 0, \label{pi0i}
\end{equation}
where $\approx$ denotes Dirac's weak equality.  With~the
momenta at hand, one thus finds the gravitational part of
the Hamiltonian (i.e., setting $L_\mathrm{m}\to 0$):
\begin{equation}
H := \int \D^3 x \( \pi^0 \dot{N} + \pi^i \dot{N}_i
+\pi^{ij} \dot{h}_{ij} \) - L\\ =  \int \D^3 x \( \pi^0
\dot{N} + \pi^i \dot{N}_i + N \Hu + N_i \Hu^i \),
\label{Hamil}
\end{equation}
where
\begin{equation}
\mathcal{H} = \frac{1}{\sqrt{h}}\left( h_{ik} h_{jl}
-\frac12 h_{ij} h_{kl}\right) \pi^{ij} \pi^{kl} -\sqrt{h}\,
\(^3\! R-2\Lambda\), \label{H}
\end{equation}
and
\begin{equation}
\mathcal{H}^i = -2\sqrt{h} \nabla_j \left(
\frac{\pi^{ij}}{\sqrt{h}} \right). \label{Hi}
\end{equation}

When adding the matter part, $\Hu$ picks a part proportional
to the energy density $\rho$, while $\Hu^i$ contains the
current $j^i$. It is then clear from \eqref{Hamil} that
Hamilton's equations imply the Hamiltonian $\Hu \approx 0$
and the diffeomorphism (momentum) $\Hu^i \approx 0$
constraints, which, together with the six dynamical
equations for $h_{ij}$, provide the canonical form of GR
that is entirely equivalent to the ten Einstein~equations.

The constraints \eqref{H} and \eqref{Hi} obey a certain
algebra which is closed but not a Lie
algebra~\cite{Kiefer:2012ria}. The~reason why this is not a
Lie algebra is the explicit occurrence of the (inverse)
three-metric in the Poisson bracket between \eqref{H}
evaluated at different space points, which in turn
originates from the constraint algebra not being the algebra
of spacetime diffeomorphisms. Generators for those
diffeomorphisms necessarily employ the lapse and the shift
functions (for~example, see Ref.~\cite{Pons:2010ad}).
For~this reason, observables in GR (and in other
diffeomorphism-invariant theories) are supposed to commute
(in the sense of Poisson brackets) with a gauge generator
that consists of a tuned {\em sum} of all constraints,
including the primary constraints \eqref{pi0i} together with
lapse and shift ~\cite{Pons:2010ad,Pitts:2016hnh}. This is
why observables in GR are not necessarily constants of
motion because~they do not need to commute separately with
the Hamiltonian and momentum constraints (as is sometimes
claimed in the literature).

\subsection{Superspace and Canonical~Quantization}

The most straightforward way to quantization once the
Hamiltonian formalism has been established consists, first
of all, in defining the relevant configuration space. In~the
case of GR, we have seen that the relevant dynamical
variables are the three-metric components $h_{ij}$ and the
matter fields jointly described by a single symbol $\Phi$.
All these depend on a coordinate parameterization on the
three-surface $\Sigma$, labeled by $\{x^a\} =: \bm{x}$. 
In~other words, we begin with the space:
$$
\mathrm{Riem}\left( \Sigma\right) := \left\{  h_{ij} \left(
x^a\right), \Phi \left( x^a \right) | \bm{x} \in \Sigma
\right\}.
$$

The space $\mathrm{Riem}\left( \Sigma\right)$ is, however,
too large as it contains many equivalent configurations,
namely those that can be related by (three-dimensional)
diffeomorphisms. Denoting by $\mathrm{Diff}\(\Sigma\)$ the
set of all such possible diffeomorphisms, one ends up taking
into account GR-invariance by reducing it to the
configuration space as the quotient:
$$
\mathrm{Conf} := \mathrm{Riem}
\(\Sigma\)/\mathrm{Diff}\(\Sigma\),
$$
an actual configuration consisting in the equivalence class
of the three-dimensional metric $h_{ij}$ and field $\Phi$. 
With~these configurations, one can consider a set of states
$\left\{ | h_{ij},\Phi \rangle \right\}$ of the space
``Conf'', called {\sl superspace}, onto~which one can
project a relevant state $ | \Psi \rangle$ to yield the
equivalent of the ``position'' representation for the wave
function, here replaced by the wave functional:
\begin{equation}
\Psi\left[ h_{ij}\(x\),\Phi\(x\)\right] := \langle
h_{ij},\Phi | \Psi \rangle. \label{WaveFunc}
\end{equation}

It is not clear whether the ensemble $\left\{ | h_{ij},\Phi
\rangle \right\}$ can actually be built explicitly or even
defined, and~the above definition is merely suggestive in
order to draw an analogy with ordinary quantum
mechanics~\cite{Hartle:1983ai}. For~a detailed discussion of
these concepts, we refer the reader to
Ref.~\cite{Giulini:2009np}.

The equations satisfied by the wave function are then
obtained by applying the Dirac quantization procedure, in
which the metric and fields are merely multiplicative
operators, and the momenta read as follows:
\begin{equation}
\pi^{ij} \to - \I\hbar \frac{\delta}{\delta h_{ij}}\ \ \ \
\hbox{and} \ \ \ \ \pi^{\Phi} \to - \I\hbar
\frac{\delta}{\delta \Phi} \label{ActMomenta}
\end{equation}
for the relevant degrees of freedom, $\pi^{0} \to - \I\hbar
\delta/\delta N$ and $\pi^{i} \to - \I\hbar \delta/\delta
N_i$ for the primary constraints, the~latter reading on~the
following state:
\begin{equation}
\hat{\pi}^0\Psi= - \I\hbar \frac{\delta\Psi}{\delta N} = 0 \
\ \ \hbox{and} \ \ \ \ \hat{\pi}^i \Psi= - \I\hbar
\frac{\delta\Psi}{\delta N_i}, \label{PiConstraints}
\end{equation}
expressing, as~anticipated in Equation~\eqref{WaveFunc},
that $\Psi$ depends neither on the lapse $N$ nor on the
shift $N_i$.\footnote{In the following we shall set
$\hbar=1$.}  With the prescription \eqref{ActMomenta}, one
transforms the momentum constraint \eqref{Hi}---now
including the matter part---into the following equation:
\begin{equation}
\hat{\Hu}^i\Psi = 0 \ \ \ \ \Longrightarrow \ \ \ \ \
\I\nabla_j^{(h)} \( \frac{\delta \Psi}{\delta h_{ij}}\) =
8\pi\GN \hat{T}^{0i} \Psi, \label{HiPsi}
\end{equation}
which can be understood to mean that configurations related
by a coordinate transformation yield similar wave
functionals; more precisely, wave functionals remain
invariant under three-dimensional diffeomorphisms connected
with the identity and acquire a phase for non-connected
(``large'') ones. Finally, the~Hamiltonian constraint
\eqref{H}---also including the matter part---yields the
equation below:
\begin{equation}
\hat{\Hu} \Psi = \left[ -16\pi\GN \mathcal{G}_{ijkl}
\frac{\delta^2}{\delta h_{ij} \delta h_{kl}} +
\frac{\sqrt{h}}{16\pi\GN}\( -^3\!R+2\Lambda + 16\pi\GN
\hat{T}^{00}\) \right]\Psi = 0, \label{HPsi}
\end{equation}
with the DeWitt metric given by the following:
$$
\mathcal{G}_{ijkl} = \frac{1}{2\sqrt{h}}\( h_{ik} h_{jl} +
h_{il} h_{jk} -h_{ij} h_{kl} \).
$$

Equation \eqref{HPsi} is called the Wheeler--DeWitt
equation. In~the way it is written here, it only has formal
significance because~we do not give a mathematical
definition for the second functional derivatives with
respect to the three-metric. This is connected to the
factor-ordering problem and requires employing a
regularization and (perhaps) renormalization. There is no
consensus on how this can be achieved, although~there are
concrete proposals in the literature~\cite{Feng:2018cul}.
Only if these technical problems are solved can we also
investigate whether an anomaly-free version of the classical
constraint algebra mentioned above can be~formulated.

The Wheeler--DeWitt Equation \eqref{HPsi} takes the form of
a stationary Schr\"odinger equation with an energy value of
zero. This points to the absence of time at the most
fundamental level and will be discussed in the next section.
An~important remark concerning the structure of~\eqref{HPsi}
is the fact that the signature of its kinetic term is
indefinite. More precisely, the~DeWitt metric can be
interpreted as a symmetric $6\times 6$ matrix (at each space
point), which can be diagonalized to finding the signature
$(-,+,+,+,+,+)$. The~minus sign is unrelated to the
Lorentzian structure of classical spacetime; it is a
consequence of the attractive nature of
gravity~\cite{Giulini:2009np,Kiefer:2012ria}. The~minus sign
is connected to the local scale $\sqrt{h}$ (``local
volume'') and gives Equation \eqref{HPsi} the structure of a
locally hyperbolic (Klein--Gordon type of) equation.
The~variable $\sqrt{h}$ connected to this sign may thus be
interpreted as an {\em intrinsic} time---time that is
entirely constructed from components of the three-metric.
Julian Barbour has emphasized the analogy of such an
intrinsic time with the ephemeris time used by astronomers
in the past ~\cite{Barbour:2009zd}.

We emphasize again that the formalism of quantum
geometrodynamics is a very conservative one.
The~Wheeler--DeWitt equation and the momentum constraints
follow in a straightforward way when we take Schr\"odingers
route of \oldstylenums{1926}, that is, ``guessing'' wave
equations that, in the classical limit, lead back to
Einstein equations in Hamilton--Jacobi~form.

\section{The Question of~Time \label{sec3}}

It is common knowledge that the physicists' view of time
evolved in \oldstylenums{1905} when Einstein made it
relative to each observer, that is, when he introduced the
notion of proper time. But~special relativistic time is
still an absolute concept as it relies on a set of
privileged inertial frames stemming from the existence of
Killing vectors in Minkowski space, which provides a fixed
(non-dynamical) background.  In~contrast, spacetime is
dynamical in~GR, and~an arbitrary spacetime does not possess
any timelike Killing vector. In~that sense, it seriously
departs from what is needed in quantum mechanics.

\subsection{Classical~Time}

A major feature of GR is its {\em background independence}.
This means that all variables are dynamical, and~absolute
structures have disappeared. In~Einstein's
words\cite{Einstein:1922}, here in J\"urgen Ehlers'
translation~\footnote{The German original reads: ``Es
widerstrebt dem wissenschaftlichen Verstande, ein Ding zu
setzen, das zwar wirkt, aber auf das nicht gewirkt werden
kann''.}~\cite{Ehlers:1995}:
\begin{quote}
  It is contrary to the scientific mode of understanding to
  postulate a thing that acts, but~which cannot be acted
  upon.
\end{quote}

This background independence is sometimes referred to as the
problem of time in the classical theory, although~in our
opinion, it is not a problem but a main, if not {\em the}
main, feature of~GR.

The canonical form of GR is very much suited to explicitly
exhibit its dynamical structure. This is achieved by
formulating ``interconnection theorems'' that demonstrate
how equations on three-dimensional hypersurfaces are
connected with spacetime equations (for~example, see
Ref.~\cite{Giulini:2006xi} and references~therein).

One theorem states the following connection: the constraints
$\Hu\approx 0$ and $\Hu^i\approx 0$ (which are four out of
the ten Einstein equations), as imposed on an ``initial
hypersurface'', are preserved in time if and only if the
energy--momentum tensor of matter has vanishing covariant
divergence. This has a nice analogy in electrodynamics: the
Gauss constraint is preserved if and only if there is charge
conservation. A~second theorem states that Einstein
equations are the unique propagation law consistent with the
constraints, that is, if~the constraints are imposed on
every hypersurface, the~connection between the hypersurfaces
must be through the remaining six dynamical Einstein
equations. Again, there is an analogy in electrodynamics:
Maxwell's equations are the unique propagation law
consistent with the Gauss~constraint.

Let us consider a classical system with $n$ variables $q^i$,
$i=1,\cdots,n$, depending on time $t$. One can formulate the
canonical equations in a parametric form by setting the
classical time variable as $t=q^0$ and adding an external
parameter $\tau$ such that $q^\mu = q^\mu(\tau)$, with~$\mu
= 0,\cdots,n$, and~$q^{\mu'}:=\D q^\mu/\D\tau$. The~action
integral then transforms into the~following:
$$
S = \int_{t_\mathrm{i}}^{t_\mathrm{f}} L\left(q^i,\dot
q^i\right) \D t = \int_{\tau_\mathrm{i}}^{\tau_\mathrm{f}}
L\left( q^\mu, \frac{q^{\mu\prime}}{q^{0\prime}}\right)
q^{0\prime} \D \tau =
\int_{\tau_\mathrm{i}}^{\tau_\mathrm{f}} \tilde{L} \,
\D\tau,
$$
where $\tilde{L}$ is a homogeneous function in
$q^{\mu\prime}$, so that
$$
\tilde{L} = \frac{\partial \tilde{L}}{\partial
q^{\mu\prime}} q^{\mu\prime} \equiv p_\mu q^{\mu\prime} \ \
\ \Longrightarrow \ \ \ \tilde{H} = p_\mu q^{\mu\prime} -
\tilde{L} =0,
$$
(with implicit summation over $\mu$), that is,
the~Hamiltonian associated with the system including time as
a variable vanishes identically. Switching back to the
original variables $t$ and its associated momentum $p_t$,
one finds that $\tilde{H}=0$ implies $H+p_t=0$, where $H$ is
the original Hamiltonian derived from the Lagrangian $L$.
Interpreting this equation as a constraint equation on the
full (unconstrained) phase space, one can reformulate it
using Dirac's weak equality: $H+p_t\approx0$. In~the next
subsection, we shall see that the quantum version of this
constraint is just the Schr\"odinger~equation.

In GR, one could hope that through a canonical
transformation, the~six geometric variables $h_{ij}$ can be
split into four embedding coordinates of the hypersurface
$\Sigma$, $\xi^\mu (x^\alpha)$ say, and two actual
gravitational degrees of freedom (which in the linearized
limit could be interpreted as the two helicity states of
weak gravitational waves). Similarly to the parametric form
above, one could hope that the vanishing Hamiltonian
\eqref{H} could be expanded to a form similar to $\Hu = p_t
+ H$, with~$H$ depending only on the relevant degrees of
freedom; in that case, the~Wheeler--DeWitt Equation
\eqref{HPsi} would exhibit a time structure similar to the
one in the Schr\"odinger equation, with~time interpreted as
an internal degree of freedom. It so happens that in some
simple cases (e.g.,~a Friedmann universe with matter content
dominated by a barotropic perfect fluid or a Bianchi~I model
in vacuum), one may identify some variable to serve as such
clocks.  Unfortunately, in~general, such a separation is not
globally feasible as time in GR is very much intertwined
with all other variables; this is called the global problem
of time~\cite{Kuchar:1991qf}. Even if such a separation can
be shown to exist in particular cases, it is hard to perform
it explicitly, and~even if it can be accomplished,
the~resulting equations can hardly be solved. This is why
this ``reduced approach'' has not proven to be a viable one
in dealing with the problem of~time.

This is different from the situation in classical mechanics.
There, one can artificially re-write the equations in a
reparametrization-invariant way. The~standard Newtonian time
$t$ then follows the demand to have the equations as simple
as possible. Or, in~the words of Henri
Poincar\'e~\cite{Poincare:1970} (our translation from
French)\footnote{The French original reads: ``Le temps doit
\^{e}tre d\'efini de telle fa\c{c}on que les \'equations de
la m\'ecanique soient aussi simples que possible. En
d'autres termes, il n'y a pas une mani\`ere de mesurer le
temps qui soit plus vraie qu' une autre; celle qui est
g\'en\'eralement adopt\'ee est seulement plus {\em
commode}''.}:
\begin{quote}
  Time must be defined in such a way that the equations of
  mechanics are as simple as possible. In~other words, there
  is no way to measure time that is more true than any
  other; the one that is usually adopted is only more {\em
  convenient}.
\end{quote}

This is not possible in GR---there is no distinguished time
that renders the equations simple in a similar~way.

\subsection{Time and the~Quantum}

As we have seen in the last subsection, treating mechanics
as a parametrized system leads to the constraint
$H+p_t\approx0$ where $H$ is the usual Hamiltonian and $p_t$
the momentum conjugate to $t$, here elevated formally to a
dynamical variable. The application of Dirac's formal
quantization rules, including $p_t\to-\I\partial/\partial t$
in particular, then leads to the Schr\"odinger equation:
\begin{equation}
 \I\frac{\partial\psi}{\partial t}=\hat{H}\psi.
 \label{Schrodinger}
\end{equation}

In the standard view of quantum mechanics, $t$ is
interpreted as an external parameter (Newton's absolute
time). That the interpretation of $t$ as an operator
(``$q$-number'') leads to problems was already recognized by
the pioneers of the theory. Wolfgang Pauli, in~his textbook
on quantum mechanics, emphasizes that the presence of an
operator $t$ obeying a canonical commutation rule with the
Hamiltonian would lead to $\hat{H}$ having a spectrum from
$-\infty$ to $+\infty$. He writes~\cite{Pauli:1990}, p.~60
(our translation from German)\footnote{The German original
reads: ``Wir schlie\ss en also, da\ss\ auf die Einf\"uhrung
eines Operators $t$ grunds\"atzlich verzichtet und die Zeit
$t$ in der Wellenmechanik notwendig als gew\"ohnliche Zahl
(`$c$-Zahl') betrachtet werden mu\ss ''.}:
  
\begin{quote}
  We thus conclude that one must completely go without the
  introduction of an operator $t$ and that the time $t$ in
  wave mechanics must necessarily be considered as an
  ordinary number (“c-number”).
\end{quote}

Erwin Schr\"odinger, in~Ref.~\cite{Schrodinger:1931},
p.~243, emphasizes (our translation from
German)\footnote{The German original reads: “Zur empirischen
Kenntnis der Zeitvariablen kann man auf keine andere Weise
als durch wirkliche Ablesung einer wirklich existierenden
Uhr gelangen. Diese Uhr ist ein physikalisches System wie
jedes andere, die Ablesung ihres Zeigerstandes eine
physikalische Messung wie jede andere. Es geht nicht an,
dieses eine physikalische System und diese eine Art von
Messungen sozusagen {\em hors concours} zu stellen und
blo\ss\ auf alle \"ubrigen die Grunds\"atze der
Quantenmechanik anzuwenden, auf die Zeitbestimmung aber
nicht”.}:
\begin{quote}
  One can arrive at an empirical knowledge of the time
  variable by no other means than by a real reading of a
  really existing clock. This clock is a physical system
  like any other, and~the reading of the pointer is a
  physical measurement like any other. It is not acceptable
  to put this particular physical system and this particular
  kind of measurements, as~we may say, {\em hors concours},
  and to apply the principles of quantum mechanics only to
  all others but~not to the determination of time.
\end{quote}

Schr\"odinger then proceeds to show that it is impossible to
construct a time operator whose eigenvalues monotonically
correlate with $t$ if one demands a bounded Hamiltonian.
In~more recent years, the conclusions by Pauli and
Schr\"odinger were re-formulated and strengthened, e.g.,~in
Ref.~\cite{Unruh:1989db}.

One can, of~course, try to re-formulate quantum mechanics by
replacing the Schr\"odinger Equation \eqref{Schrodinger}
with an equation that does not refer to the external $t$,
but~to a variable describing a real
clock~\cite{Malkiewicz:2017cuw}. Such an approach is
realized in the ``Montevideo interpretation'' of quantum
theory ~\cite{Gambini:2020bup}. One finds in this way a
master equation of the Lindblad type for an effective
density matrix, which does not evolve unitarily.

Schr\"odinger claimed that this conceptual problem can only
be solved by developing a relativistic framework. As~we have
seen, the~background independence of GR brings this
discussion to a new level and leads, after~quantization,
to~the quantum version of the problem of time, to~which we
now~turn.

\subsection{The~Problem}

\begin{figure}[H]
\centering
\includegraphics[width=12cm]{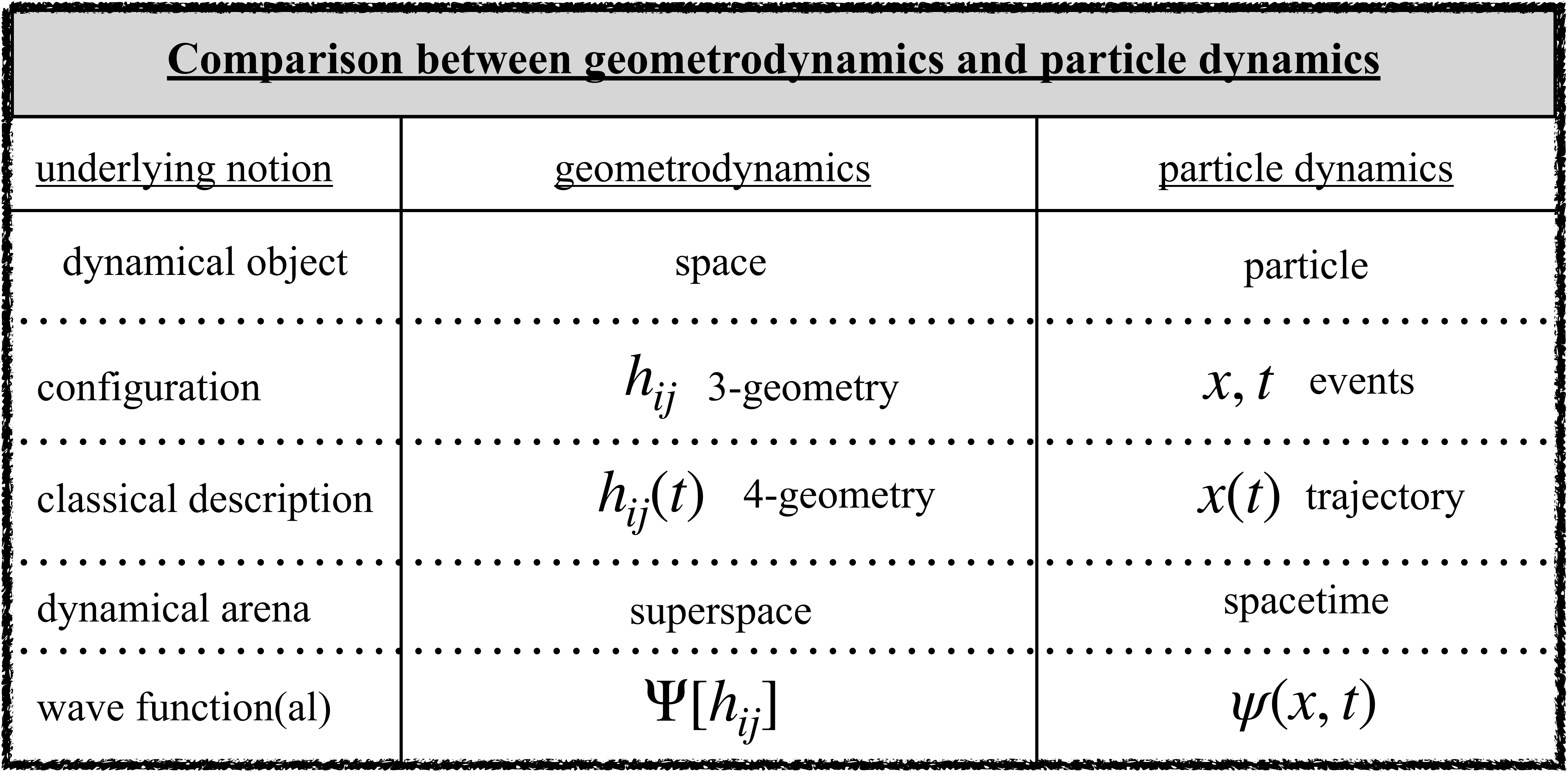}

\caption{Geometrodynamics shares many properties with
particle dynamics, and the~relevant notions in both can be
compared as shown. Modified and extended version of Box 43.1
in Ref.\cite{Misner:1973prb}.}

\label{MTW_Box}
\end{figure}

We have seen that one can understand GR as that which
provides (generalized) trajectories of three-dimensional
space, in much the same way as usual classical mechanics
provides particle trajectories. Spacetime can be interpreted
as a class of trajectories in the space $\mathrm{Riem}\left(
\Sigma\right)$.  Since the classical particle trajectories
are absent in quantum mechanics, one expects that the same
applies to spacetime in quantum gravity. In~fact, this is
what the application of the standard quantization rules, as
reviewed above, gives. Figure~\ref{MTW_Box}, which is a
modified and extended version of Box 43.1 in
Ref.~\cite{Misner:1973prb}, presents a comparison of
geometrodynamics with particle dynamics, from~which it
becomes clear that the quantum states can be represented by
wave functionals on the space of {\em three}-geometries, not
{\em four}-geometries.

Therefore, at the fundamental level, spacetime (and with it,
time) has disappeared, and~only space remains. This
consequence holds for any theory that is
reparametrization-invariant at the classical level. As
mentioned above, it is generally not possible to
consistently rewrite the constraint equations of GR in a
form similar to $H+p_t\approx0$. All momenta in the
Wheeler--DeWitt Equation \eqref{HPsi} appear quadratically,
and~there is no distinguished part of the momenta that may
serve as being conjugate to a distinguished time. In~this
sense, the~classical ``problem of time'' (background
independence) leaves its imprint on quantum~theory.

The discussion presented here is based on the canonical
(Hamiltonian) formalism of quantum GR. Alternatively, one
can use the covariant (path-integral) formulation. At~the
formal level, that is, neglecting field-theoretic
subtleties, the~two formulations are equivalent: the path
integral satisfies the Wheeler--DeWitt equation and the
momentum constraints~(\cite{Kiefer:2012ria}, Section~5.3.4).
For~this reason, all issues related to the problem of time
hold equally well in the covariant~formulation.

In quantum mechanics, the~evolution with respect to the
external time $t$ is unitary for~closed systems, that is,
probabilities are conserved. If~time is absent, the~question
arises whether the concepts of probability and unitarity
still make sense. These issues are always included, at~least
implicitly, when addressing the problem of~time.

The absence of spacetime, and with it the absence of any
external time variable at the most fundamental level, is
usually what is meant when one talks about the problem of
time in quantum gravity and quantum cosmology. What are the
suggestions for its solution?

\subsection{Two~Solutions}

We present here two main solutions to address the problem of
time. For~a comprehensive review on potential solutions, we
recommend Ref.~\cite{Anderson:2017jij} and the 933
references~therein.

\subsubsection{Intrinsic~Time} \label{sec_Intrisic}

The most straightforward approach is to accept the absence
of spacetime at the fundamental level and to search for an
interpretation that is solely based on three-metrics (and
-geometries). The~standard concept of time then only emerges
at an appropriate semiclassical limit (see
Section~\ref{sec_semclas} below). The~semiclassical limit is
in any case demanded for consistency and is all that is
presently available for experimental or observational~tests.

As we have already mentioned above, the~Wheeler--DeWitt
Equation \eqref{HPsi} is of a locally hyperbolic form,
with~the local scale $\sqrt{h}$ coming with a different sign
in the kinetic term. One could then call the local scale a
local {\em intrinsic time}. Intrinsic time is entirely
constructed from spatial degrees of freedom (e.g., see
Ref.~\cite{Malkiewicz:2019azw}).

Most discussions in quantum cosmology deal, in~fact,
with~simplified models that assume a model of a homogeneous
universe. Since most degrees of freedom of the full theory
are absent then (there are, in~particular, no gravitational
waves), one calls them {\em minisuperspace}
models~\cite{Misner:1972}. For simplicity, let us choose a
closed Friedmann--Lema\^{\i}tre universe with scale factor
$a$, containing a homogeneous massive scalar degree of
freedom $\phi$; this gives a two-dimensional configuration
space. Classically, the line element reads as follows:
\begin{equation}
\D s^2=-N^2(t)\D t^2+a^2(t)\D\Omega_3^2,
\label{miniRW}
\end{equation}
where $\D\Omega_3^2$ is the standard line element on the
three-sphere. The quantum version invokes a wave function
depending on the two variables, $\psi(a,\phi)$. The~momentum
constraints are identically fulfilled, and~the
Wheeler--DeWitt equation reads as follows (in units where
$2\GN/3\pi=1$):
\begin{equation}
  \label{wdw-mini}
\frac{1}{2}\[\frac{1}{a^2}\frac{\partial}{\partial a}
\left(a\frac{\partial}{\partial a}\right)-\frac{1}{a^3}
\frac{\partial^2}{\partial\phi^2}-a+\frac{\Lambda a^3}{3}+
m^2a^3\phi^2\]\psi(a,\phi)=0.
\end{equation}

Here, the~factor ordering is chosen in order to achieve
covariance in minisuperspace.

The hyperbolic structure is obvious from \eqref{wdw-mini}:
the kinetic term for the scale factor has a different sign.
That this sign is connected with $a$ and not with $\phi$ can
be recognized by adding further degrees of freedom,
for~example, shape degrees and additional matter~fields.

The structure of \eqref{wdw-mini} gives rise to two
drastically different types of determinism in the classical
and quantum theory~\cite{Zeh:2007,Kiefer:2012ria} (see
Fig~\ref{Fig_aphi} below, taken from~\cite{Kiefer:2012ria}).
In~the classical theory, we have a trajectory in the
$(a,\phi)$-configuration space found by imposing initial
conditions at one side of the trajectory (e.g., near the
``big bang''). For~a recollapsing universe, the~collapsing
part of the trajectory can be considered as the
deterministic successor of the expanding part; ``big
crunch'' is different from ``big~bang''.

\begin{figure}[h]
\begin{minipage}[t]{0.4\textwidth}
\begin{center}
   {\small\bf Classical theory}\\[5mm]
\includegraphics[width=0.9\textwidth]{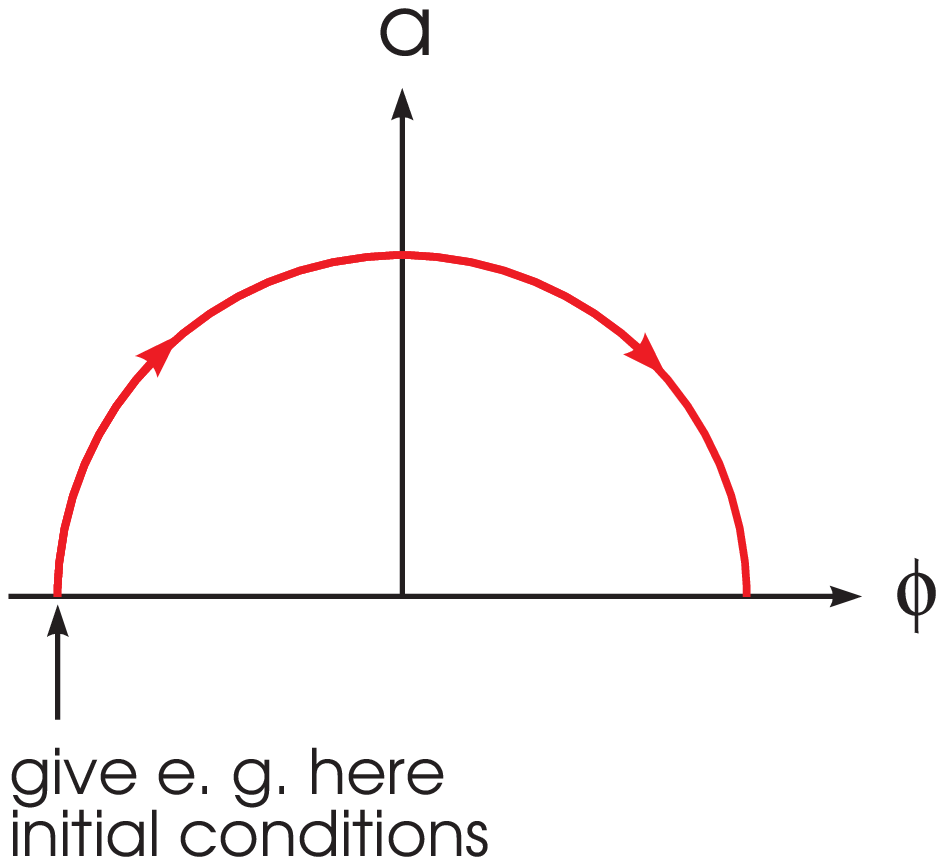}\\[5mm]
{\small Recollapsing part is\\ deterministic successor~of\\
expanding part}
\end{center}
\end{minipage}\hfill
\begin{minipage}[t]{0.49\textwidth} 
\begin{center}
   {\small\bf Quantum theory}\\[5mm]
\includegraphics[width=0.75\textwidth]{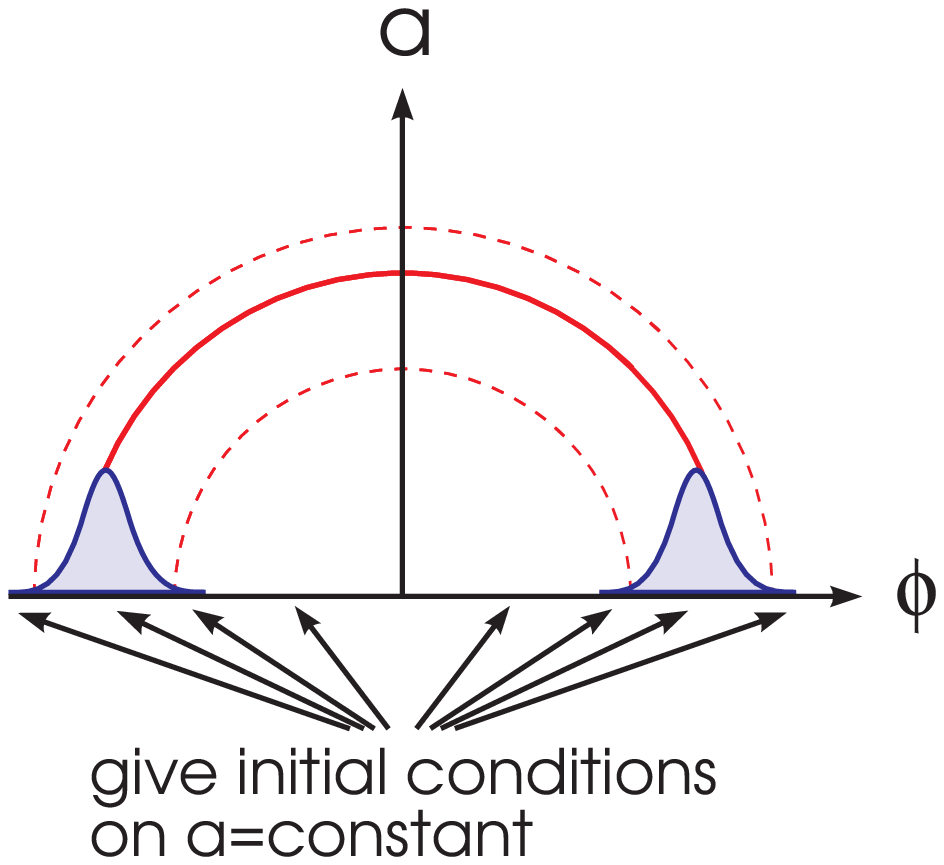}\\[5mm]
{\small `Recollapsing' wave packet must be present `initially'}
\end{center}
  \end{minipage}
\caption{Comparison between classical trajectory and the
quantum wavefunction for a classically expanding and
collapsing universe.}
\label{Fig_aphi}
\end{figure}

It is not so in quantum theory. For the wave Equation
\eqref{wdw-mini}, initial conditions must be formulated at
constant $a$. In~order to construct a wave packet following
the classical trajectory as a narrow tube, the~initial
condition at $a=$ constant has to include both the ``big
bang'' piece {\em and} the ``big crunch'' piece; in
configuration space, they are both part of $a=$ constant.
The~quantum determinism acts from small $a$ to large $a$,
not along a classical trajectory (which is absent).

This new type of determinism in quantum cosmology has
important consequences when discussing the origin of the
arrow of time~\cite{Zeh:2007}. For~an initial condition of
small entropy at small scale factor, the~arrow of time would
formally reverse at the classical turning
point~\cite{Kiefer:1994gp}. The~reason for this is that
small entropy at small $a$ would refer to both the ``big
bang'' piece and the ``big crunch'' piece of the wave
packet; entropy would increase with increasing $a$ and not
along a classical trajectory (classical spacetime). A~low
initial entropy could come from a quantum version of
Penrose's Weyl curvature hypothesis~\cite{Kiefer:2021zqz}.
Quantum geometrodynamics is able to implement the avoidance
of cosmic singularities by invoking DeWitt's
criterion~\cite{DeWitt:1967yk}: regions in configuration
space are avoided in quantum gravity if the wave function
vanishes there. This criterion can be generalized to
accommodate the conformal structure of the configuration
space~\cite{Kiefer:2019bxk}. With~this, singularity
avoidance has been shown for isotropic and anisotropic
models (for~example, see
Refs.~\cite{Kiefer:2019bxk,Chataignier:2019kof} and the
references therein). These insights demonstrate the
importance of a thorough understanding of the concept of
time in quantum~cosmology.

As is well known, there is still an ongoing debate about the
interpretation of quantum theory. The~situation becomes
especially demanding in quantum cosmology, where by
definition we are dealing with a quantum theory of a closed
system, with~no reference to external observers or
measurement agencies. Accepting the linear nature of quantum
theory and the universal validity of the superposition
principle, one natural way would be to invoke the Everett or
``many worlds'' interpretation.\footnote{The name ``many
worlds'' may, strictly speaking, be inappropriate because
one deals with one {\em quantum} world. In~fact, Everett
himself used the term ``relative states''.} Bryce DeWitt,
in~his pioneering paper on canonical quantum theory,
writes~\cite{DeWitt:1967yk}:
\begin{quote}
   Everett's view of the world is a very natural one to
   adopt the quantum theory of gravity, where one is
   accustomed to speak without embarassment of the `wave
   function of the universe.' It is possible that Everett's
   view is not only natural but essential.
\end{quote}

It has indeed become possible to speak without embarassment
of the wave function of the universe~\cite{Hartle:1983ai}. 
It is no coincidence that Everett's interpretation became
more wildly known only after DeWitt had made use of it in
quantum cosmology. In~the next subsection, we discuss a
different approach to interpreting quantum cosmology and the
problem of~time.

\subsubsection{The Trajectory~Approach}

As one way to understand the problem of time consists in
arguing that one cannot reconstruct a four-dimensional
spacetime from a given three-dimensional initial
hypersurface because of the lack of dynamical equations, it
is natural to think that having such dynamical equations at
the quantum level could actually solve the problem (note
that it can also lead to new effects that could potentially
be observed in a cosmological
setup~\cite{Valentini:2021izg}).  In~the case of the
Schr\"odinger equation, this implies particle trajectories,
although~not the classical ones, as~suggested in
\oldstylenums{1927} by de Broglie~\cite{deBroglie:1927} and
further completed 25 years later by
Bohm~\cite{Bohm:1951xw,Bohm:1951xx}. Similarly, actual
functions $h_{ij}(t)$ could be~obtained.

In quantum mechanics, this works as follows. By making the
amplitude and phase of the wave function explicit through
$\psi(\bm{x},t) = |\psi| \E^{\I S}$ and setting $\hat{H} = -
\bm{\nabla}^2/(2m) +V(\bm{x})$ in the Schr\"odinger Equation
\eqref{Schrodinger}, one gets the following continuity
equation:
\begin{equation}
\frac{\partial |\psi|^2}{\partial t} + \bnab \cdot \(
|\psi|^2 \frac{\bnab S}{m} \) =0, \label{contdBB}
\end{equation}
and a modified Hamilton--Jacobi equation:
\begin{equation}
\frac{\partial S}{\partial t} + \frac{\( \bnab S\)^2}{2 m} +
V(\bm{x}) -\frac{\bnab^2|\psi|}{2m|\psi|} =0, \label{HJdBB}
\end{equation}
by splitting \eqref{Schrodinger} into its real and imaginary
parts. Equation \eqref{contdBB} leads to the usual Born
probability density $\rho = |\psi|^2$, while the momentum
$\bm{p}$ can be identified with $\bm{\nabla}S/m$; the
classical potential is then seen to be corrected by a
quantum potential $Q$ given by the following:
\begin{equation}
Q(\bm{x},t) = - \frac{\bnab^2|\psi|}{2m|\psi|} = -\frac{1}{4
m \rho}
\[ \bnab^2 \rho - \frac{\(\bnab \rho\)^2}{2\rho} \]. 
\label{QdBB}
\end{equation}

Assuming now an actual trajectory $\bm{x}(t)$, the~momentum
can be written as $\bm{p} = m\bm{v} = m \dot{\bm{x}}$,
thereby defining a velocity $\bm{v}$ that satisfies the
following guidance equation:
\begin{equation}
\frac{\D\bm{x}}{\D t} = \bnab S \qquad \Longrightarrow
\qquad
m\frac{\D^2\bm{x}}{\D t^2} = - \bnab \[  V(\bm{x}) + Q(\bm{x},t) \],
\label{vdBB}
\end{equation}
which induces a quantum correction to the classical
evolution. Note that it also naturally provides a clear
distinction between the classical and the quantum regimes,
the~former being recovered in the limit $Q\to 0$, a
well-defined and state-dependent~statement.

Let us restrict our attention to a minisuperspace model, in
which one formally replaces $h_{ij}$ with a simpler
three-metric $h_{ij}^{\!\text{(sym)}}$, including symmetries
(e.g., homogeneity and/or isotropy) and~matter content in a
set $\{ q_a \}$ standing collectively for the geometric
degrees of freedom that are relevant to encode the
symmetries and the variables that describe matter; in the
case of \eqref{miniRW}, for instance, one would have $\{ q_a
\} = \{ a, \phi \}$. In~terms of these variables, the~DeWitt
metric is reduced from $G_{ijkl} (h_{ab})$ to $\Gamma_{ab}
(q_c)$, and~writing the last term in \eqref{HPsi} containing
the intrinsic scalar curvature $^3R$ and the $00$ component
of the stress tensor $\hat{T}^{00}$ as a mere function $V$
of the $q_a$, the~Wheeler--DeWitt equation
becomes~\cite{Pinto-Neto:2018zvn}:
\begin{equation}
- \frac12 \Gamma_{ab} \frac{\partial^2 \Psi (q_c)}{\partial
q_a \partial q_b} + V(q_c)  \Psi (q_c) = 0. \label{miniWDW}
\end{equation}

Equation \eqref{miniWDW} is much simpler than \eqref{HPsi}
and can in fact be solved exactly in many~cases.

By expanding the wave functional into amplitude $|\Psi|$ and
phase $S$ once again, one gets the modified Hamilton--Jacobi
equation: \vspace{6pt}
\begin{equation}
\frac12 \Gamma_{ab} \frac{\partial S}{\partial q_a}
\frac{\partial S}{\partial q_b} + V(q_c) \underbrace{-
\Gamma_{ab} \frac{\partial^2 |\Psi|}{|\Psi| \partial q_a
\partial q_b}}_{Q(q_q)} = 0, \label{modHJWDW}
\end{equation}
whose solution yields the function $S(q_a)$, and
\begin{equation}
\Gamma_{ab} \frac{\partial}{\partial q_a} \( |\Psi|^2
\frac{\partial S}{\partial q_b} \) = 0, \label{contWDW}
\end{equation}
hence, one can identify the ``velocity'' $\D q_a/(N\D t)$
through
\begin{equation}
\frac{\partial S}{\partial q_a} = \Gamma^{ab} \frac{\D
q_b}{N\D t}. \label{GuiWDWdBB}
\end{equation}
This gives $q_a(t)$, that is, a~full reconstruction of a
four-dimensional spacetime. Note that with Equation
\eqref{GuiWDWdBB}, being invariant under time
reparametrization, the~resulting spacetime geometry is
independent of the choice of the lapse function $N$,
as~required.

In general, however, the~trajectory approach cannot really
come to an absolute conclusion as to whether there exists a
satisfying solution to the problem of time in quantum
gravity, even though it is the case for at least the quantum
cosmological setup in which one restricts attention to
minisuperspace. We refer the reader to
Ref.~\cite{Pinto-Neto:2018zvn} and the references given
therein, in~which a detailed discussion of the possible wave
functional configurations and their consequences
is~presented.

\subsection{Time from Semiclassical~Gravity}
\label{sec_semclas}

Since the early days of quantum theory, scientists have
already speculated about the meaning of the $t$ in the
Schr\"odinger Equation \eqref{Schrodinger}. We have already
referred to the work of Pauli and Schr\"odinger.
In~\oldstylenums{1931}, Neville Mott made another important
contribution~\cite{Mott:1931}.  He proposed that it is not
the time-dependent but~the time-{\em independent}
Schr\"odinger equation, $H\psi=E\psi$, that is fundamental.
The~time-dependent version emerges from a {\em correlation}
between subsystems in the total time-independent~system.

As a concrete example, Mott considered an electron,
described by position ${\bm r}$, interacting with an
alpha-particle, described by position ${\bm
R}\equiv(X,Y,Z)$.  By making the following ansatz:
\begin{equation}
\Psi({\bm r},{\bm R})=f({\bm r},{\bm R})\E^{\I
  kZ},
\end{equation}
he was able to {\em derive} an effective time-dependent
Schr\"odinger equation for the electron, with~time $t$
constructed from the $Z$-coordinate of the alpha-particle.
In~this, he explicitly referred to the approximation
introduced by Born and Oppenheimer in \oldstylenums{1927}.
Time is thus defined by the alpha-particle. The~general idea
of deriving evolution entirely from internal clock readings
was elaborated on in detail by Ref.~\cite{Page:1983uc}.

A similar mechanism is at work in quantum cosmology. There,
a~semiclassical time can emerge from the timeless
Wheeler--DeWitt equation by dividing the total system into
subsystems, with one of these subsystems capable of
providing a time variable in reference to which the other
subsystems evolve. To~be more concrete, the~limit of quantum
field theory in an external spacetime can be obtained by a
Born--Oppenheimer type of expansion scheme with respect to
the Planck-mass squared, $m_\textsc{p}^2$
\cite{Kiefer:1990pt}. One starts with the following ansatz
for the total wave function(al) of gravity and matter:
$$
\Psi[h_{ab},\phi]\equiv\E^{\I S[h_{ab},\phi] }
$$
and expands the exponent in powers of $m_\textsc{p}^2$,
$$
S[h_{ab},\phi]=m_\textsc{p}^2S_0+S_1+m_\textsc{p}^{-2}S_2+\ldots .
$$

This is then inserted into the full Wheeler--DeWitt
equation, and~different powers of $m_\textsc{p}^2$ are
compared. The~highest orders, $m_\textsc{p}^4$ and
$m_\textsc{p}^2$, lead to a $\phi$-independent $S_0$, for
which the gravitational Hamilton--Jacobi equation holds.
In~this way, (vacuum) GR is recovered.\footnote{Non-vacuum
GR can be recovered by adding some of the $\phi$-degrees of
freedom to $S_0$.}

At the order $m_\textsc{p}^0$, one obtains an equation for
$S_1$ that can be simplified by introducing the following
equation:
\begin{equation}
\psi:= D[{h}_{ij}] \E^{\I S_{1}}
\end{equation}
and demanding the standard ``WKB prefactor equation'' for
$D$, which is actually similar to~\eqref{contWDW}. Next, one
can define a local ``bubble'' (Tomonaga--Schwinger) time
functional by:
\begin{equation}
\frac{\delta}{\delta\tau(\X)}:= G_{abcd}\frac{\delta
S_0}{\delta h_{ab}(\X)} \frac{\delta}{\delta h_{cd}(\X)};
\end{equation}
with this, we obtain the following Tomonaga--Schwinger
(local Schr\"odinger) equation:
\begin{equation}
    \I\frac{\delta}{\delta\tau(\X)}\psi=\mathcal{H}_{\rm
    m}(\X)\psi, \label{TS}
\end{equation}
where $\mathcal{H}_{\rm m}(\X)$ is the matter Hamiltonian
density.  We note that $\tau(\X)$ is not a scalar function
because~the commutator of $H_{\rm m}$ at different space
points does not vanish~\cite{Giulini:1994cz}.  Nevertheless,
one can integrate \eqref{TS} over $\D^3x$ with a chosen
lapse function to yield a (functional) Schr\"odinger
equation with respect to a ``WKB time'' $t$ and the
non-gravitational part of the full Hamiltonian.  This
equation describes the limit of quantum field theory in
curved spacetime. The~next order, $m_\textsc{p}^{-2}$, gives
genuine quantum-gravitational correction terms to the
matter~Hamiltonian.

There has been a debate about whether the Schr\"odinger
equation with the quantum-gravitational corrections evolves
unitarily or not with respect to WKB time. Since this
equation is derived from the Wheeler--DeWitt equation, one
would not expect that it does. One {\em can}, however,
construct a physical inner product by an intricate
construction in reference to which there is a unitary
evolution~\cite{Chataignier:2019kof}. Using these correction
terms, one can unambiguously calculate quantum-gravitational
corrections to the CMB power
spectrum~\cite{Chataignier:2020fap}.  These are observable
in principle but~not in practice because~the factor
$m_\textsc{p}^{2}$ in the denominator makes them too~small.

The derivation of the Schr\"odinger equation from the
timeless Wheeler--DeWitt Equation \eqref{HPsi} works only
for complex solutions of the form (at the highest order)
$\exp(\I m_\textsc{p}^2S_0)$, with~a real $S_0$. The~$\I$
there is the same $\I$ that occurs in the Schr\"odinger
equation with respect to WKB time. It is only for such
complex states that standard quantum (field) theory can be
derived as an approximation from timeless quantum
gravity.\footnote{A general discussion of why we need
complex quantum states and where they come from can be found
in the twin papers~\cite{Barbour:1993iv,Kiefer:1993yg} and
the references therein.} For such states, one can invoke the
standard notions of probability and unitarity. In~light of
the full Wheeler--DeWitt equation, these are only
approximate (``phenomenological'', in~the words of DeWitt)
notions. Such states occur in the BO-approximation discussed
above, but~can also be introduced beyond
it~\cite{Chataignier:2019kof}. One can also argue from the
trajectory approach that the probability interpretation in
the form of the Born rule can only emerge in the
semiclassical limit~\cite{Valentini:2021izg}.

General states can be found in these complex states by the
application of the superposition principle.
Superposing,~particularly $\exp(\I m_\textsc{p}^2S_0)$ with
its complex conjugate, one obtains a real (approximate)
solution to the Wheeler--DeWitt equation. The~notions of WKB
time and unitarity can only be employed for the two complex
components in this superposition provided these components
{\em decohere}. Decoherence comes from the interaction
between relevant and irrelevant degrees of freedom and
guarantees that in most (but not all) situations usually
dealt with in cosmology, the~classical appearance of (most)
gravitational degrees of freedom holds (for~example, see
Ref.~\cite{Kiefer:2012ria}, Chap.~10, and~the references
therein).  The~time concept discussed in
Section~\ref{sec_Intrisic} above, together with the recovery
of time from semiclassical gravity discussed here, provides
a minimal solution to the problem of time and is in
accordance with all observations made so~far.

Some final remarks are in order for isolated
quantum-gravitational systems such as (quantum) black holes.
For~such systems, one can derive an effective Schr\"odinger
equation in which $t$ is the WKB time of the (semiclassical)
Universe in which the black holes are embedded, but~where
the Hamiltonian is the Wheeler--DeWitt Hamiltonian for the
quantum black hole~\cite{Kiefer:2009zwx}. For~such isolated
systems, therefore, there is no problem of time, and~the
standard notions of probability and unitarity apply.
The~problem of time is a problem for closed
quantum-gravitational systems such as the Universe as
a~whole.

\section{Conclusions \label{sec4}}

The question of time in quantum cosmology represents a
serious challenge to be understood fully in order to build a
complete quantum gravity theory from which the observed
world, based on the classical theory of general relativity,
can be derived. The~ability to produce a consistent
four-dimensional spacetime must indeed be ensured, a~highly
nontrivial task within all the known setups or
approximations that have been proposed to implement its
quantization. Here, we have discussed the case of canonical
GR through the Wheeler--DeWitt equation, and we have shown
but a few plausible solutions among
many~\cite{Anderson:2017jij,Oriti:2021zju}.

In spite of this restriction, we believe that aspects of the
problem of time discussed in this article as~well as their
possible solutions can be found for a large class of
quantum-gravity approaches, definitely for those approaches
that arise heuristically from a formal quantization of a
classically diffeomorphism-invariant theory. This is evident
for all generalized geometrodynamic theories of gravity,
such as $f(R)$-theories, which can be re-formulated as
Einstein gravity plus a scalar field. An~explicitly
discussed example is conformal (Weyl) gravity where a
concept of shape time emerges ~\cite{Kiefer:2016uqm}.
The~situation is somewhat different in loop quantum gravity,
which is a theory that follows from a quantization of GR,
but~which makes use of different variables. For~this reason,
the~concept of time exhibits subtleties in addition to the
ones discussed here for geometrodynamics
(see~\cite{Rovelli:2018xqw} and the references therein). 
The~situation is really different in string theory,
because~this is not a direct quantization of GR but~a
fundamental quantum theory of all interactions from where
quantum gravity arises as an {\em emergent} theory.
However,~it is also claimed there that spacetime is not
fundamental; instead, it must be constructed from a
holographic, dual theory~\cite{Horowitz:2004rn}. A~detailed
discussion of time in these more generalized theories is
beyond the scope of our~paper. \vspace{6pt}

\authorcontributions{Writing---review and editing, C.K. and
P.P.  All authors have read and agreed to the published
version of the manuscript.}

\funding{This research received no external~funding.}

\institutionalreview{ Not applicable.}

\informedconsent{ Not applicable.}

\dataavailability{ Not applicable.} 

\conflictsofinterest{The authors declare no conflicts of~interest.} 

\begin{adjustwidth}{-\extralength}{0cm}
\printendnotes[custom]

\reftitle{References}

\end{adjustwidth}

\end{document}